\documentclass[apjl]{emulateapj}
\usepackage{graphics,epsf}
\usepackage{amsmath}                
\usepackage{amsfonts}               
\usepackage{amssymb}                
\usepackage{epsfig}                 

\def \K{~\rm{K}}

\def \AU{~\rm{AU}}
\def \erg{~\rm{erg}}
\def \yrs{~\rm{yrs}}
\def \yr{~\rm{yr}}

\def \days{~\rm{days}}

\begin{document}

\title{PERIASTRON PASSAGE TRIGGERING OF THE 19TH CENTURY ERUPTIONS OF ETA CARINAE}

\author{Amit Kashi\altaffilmark{1} and Noam Soker\altaffilmark{1}}

\altaffiltext{1}{Department of Physics, Technion $-$ Israel Institute of
Technology, Haifa 32000 Israel; kashia@physics.technion.ac.il;
soker@physics.technion.ac.il.}

\begin{abstract}
We reconstruct the evolution of $\eta$ Car in the last two centuries,
under the assumption that the two 19th century eruptions were triggered
by periastron passages, and by that constrain the binary parameters.
The beginning of the Lesser Eruption (LE) at the end of the 19th century
occurred when the system was very close to periastron passage,
suggesting that the secondary triggered the LE.
We assume that the 1838-1858 Great Eruption (GE) was triggered by
a periastron passage as well.
We also assume that mass transferred from the primary to the secondary
star accounts for the extra energy of the GE.
With these assumptions we constrain the total mass of the binary
system to be $M=M_1+M_2 \ga 250 \rm{M_\odot}$.
These higher than commonly used masses better match the observed
luminosity with stellar evolutionary tracks.
Including mass loss by the two stars and mass transfer from the primary to
the secondary we obtain a good match of periastron passages to the two peaks
in the light curve of the GE.
Based on these finding and a similar behavior of P~Cygni,
we speculate that major LBV eruptions are triggered by stellar companions,
and that in extreme cases a short duration event with a huge mass transfer
rate can lead to a bright transient event on time scales of weeks to months
(a ``supernova impostor'').
\end{abstract}

\keywords{ (stars:) binaries: general$-$stars: mass loss$-$stars:
winds, outflows$-$stars: individual ($\eta$ Car)}

\section{INTRODUCTION}
\label{sec:intro}

\begin{figure}[t]
\resizebox{0.5\textwidth}{!}{\includegraphics{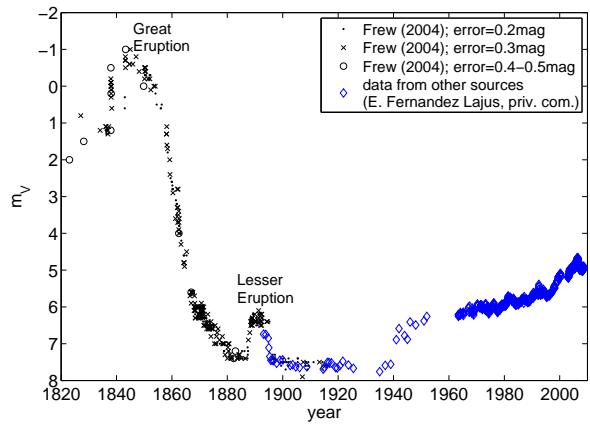}}
\caption{\footnotesize
The historical V-mag light curve
(Frew 2004, marked by black symbols; Humphreys et al. 1999 \& Fernandez Lajus, private communication, marked by blue diamonds)
during the last two centuries.
The two major eruptions, the 1837.9 $-$ $\sim$1858 Great Eruption (GE),
and the 1887.3-1895.3 Lesser Eruption (LE) are marked on the figure.
The estimated errors of the observations of Frew (2004) are given in the legend.
The two rapid rises in V-mag in 1837.9 and 1843 at the beginning of the GE are clearly seen.
Even though that the error in the peaks is 0.4-0.5 mag, it is evident
that the rise in these two peaks was $\ga 1.5~$mag.
}
\label{fig:historic_eta_car}
\end{figure}

$\eta$ Car is a binary system (Damineli 1996) containing a very massive Luminous Blue Variable
(LBV; hereafter `the primary') and a hotter and less luminous evolved main sequence companion
(hereafter `the secondary').
The LBV has undergone two major eruptions in the 19th century.
The Great Eruption (GE) took place between 1837.9 $-$ $\sim$1858 and created
the bipolar Homunculus nebula which contains $10-40 \rm{M_\odot}$, and possibly more.
(Gomez et al. 2006, 2009; Smith et al. 2003; Smith \& Ferland 2007; Smith \& Owocki 2006).
Smith (2009) has even suggested that the Homunculus can be more massive.
Following the GE, the Lesser Eruption (LE) took place between 1887.3-1895.3.
This was a much less energetic eruption (Humphreys et al. 1999)
and only $0.1-1 \rm{M_\odot}$ were ejected from the primary (Smith 2005).
A summary of the observed visual magnitude from the earliest measurements
in the 17th century, through the 19th century eruptions and
up to 2004 was prepared by Frew (2004).
Presently $\eta$ Car is being continually monitored in the optical wavebands by
the La Plata observatory (Fernandez Lajus et al. 2010),
and occasionally by other observatories.
Figure \ref{fig:historic_eta_car} presents the visible observations from the last
two centuries.

\begin{figure}[!t]
\resizebox{0.50\textwidth}{!}{\includegraphics{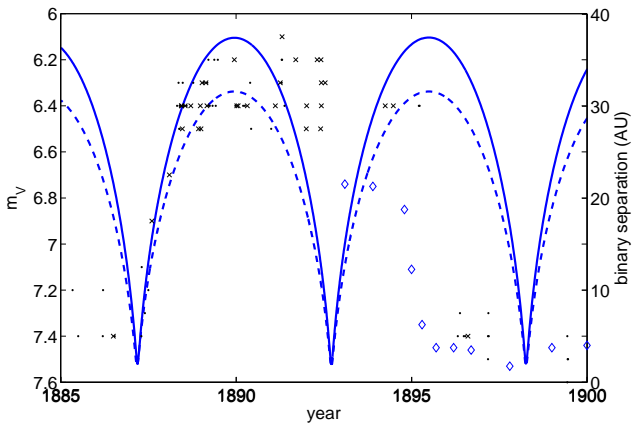}}
\caption{\footnotesize
The binary separation (blue lines) and the V-mag light curve
(Frew 2004, marked by black symbols, see legend in Fig. \ref{fig:historic_eta_car};
Humphreys et al. 1999 \& Fernandez Lajus, private communication, marked by diamonds)
during the Lesser Eruption (LE) of $\eta$ Car (1887.3-1895.3).
Dashed blue line: counting periastron passages back in time,
not taking into account any mass loss or mass transfer.
We used the common parameters $P_0=5.539\yrs$, $a_0=16.64\AU$, $e_0=0.9$,
$M_1=120 \rm{M_\odot}$ and $M_2=30 \rm{M_\odot}$
to draw the binary separation, but individually these parameters, apart from $P_0$, have no effect on
the times of periastron passages.
It is evident that the LE started very close to periastron (Frew 2004).
Solid blue line: including mass loss (for a total of $1\rm{M_\odot}$)
and mass transfer from the primary to the secondary (for a total of $0.14 \rm{M_\odot}$) at a
constant rate for the duration of the LE.
The masses before the event are $M_1 = 170 \rm{M_\odot}$ and $M_2=80 \rm{M_\odot}$,
to match the values at the end of the GE in the model named `MTz' described in section \ref{subsec:results_basic}.
We assume that during the periastron passage the interaction of an already
unstable primary star with the secondary triggered the LE.
}
\label{fig:LE_plot2}
\end{figure}

At present the orbital period is $P_0=2023 \days = 5.539 \yrs$
(Damineli et al. 2008a; Fernandez Lajus et al. 2010; Landes \& Fitzgerald 2009),
and the orbit is very eccentric.
We will use an eccentricity of $e_0=0.9$ as in our previous papers
(e.g., Kashi \& Soker 2009).
Counting back in time the periastron passages, simply by subtracting a natural
multiplicity of $P_0$, it is evident that the LE started very close to periastron (Frew 2004).
This is shown by the dashed blue line in Fig. \ref{fig:LE_plot2}.
The beginning of the LE just as periastron passage occurred,
leads us to our basic assumption:
The \emph{Major} LBV eruptions of $\eta$ Car are triggered by the periastron passages of the secondary.
The major eruptions are defined as eruptions in which the luminosity increases by a few magnitudes,
as oppose to regular eruptions, e.g. S Dor phases, or weak eruptions on top of the GE light curve,
in which the luminosity is changed by $\lesssim 0.5$ mag.
The primary must already be in a very unstable phase in order for the secondary to trigger its major eruption.
For that, such a major eruption does not occur at each periastron passage.
It is the companion that makes the eruption so energetic.

We shall use the common definition that phase zero is defined at periastron passage.
Note that the so called spectroscopic event in present $\eta$ Car $-$ the weakening or even disappearance of ionization lines
and the minimum intensity in all bands $-$ does not necessary occur exactly at periastron,
but possibly a few days earlier or later (Damineli et al. 2008a,b).
This has little importance for our model, in which we only interested when the periastron passages occur,
and not in the spectroscopic event.
We only assume that periastron took place a few days earlier or later than Jan 11, 2009,
as observed (Fernandez Lajus et al. 2010; Landes \& Fitzgerald 2009).
As will be evident below, for our calculation
the possible few days error in determining the exact time of present periastron passages presently,
is much smaller than the other sources of error.

Previously, under different considerations, Livio \& Pringle (1998) speculated that the GE
was triggered by strong interactions among the stars and their winds at periastron passage.
Soker (2001, 2004, 2005, 2007) suggested that the GE was caused by
disturbances in the outer boundary of the inner convective region,
which expelled the outer radiative zone.
According to these papers, this resulted in a mass loss of $\sim 20 \rm{M_\odot}$ from the primary star,
of which $12 \rm{M_\odot}$ were accreted onto
the secondary that ejected $4 \rm{M_\odot}$ of them by
blowing bipolar jets, and kept the other $8 \rm{M_\odot}$.
The accretion rate was high, and the potential lobe of the primary is probably filled for most of the orbit
(the potential lobe is the Roche lobe analogue,
as here there is no synchronization while the RLOF process assume synchronization).
In addition Bondi-Hoyle accretion (wind accretion) was also present.
We studied these two processes in the context of the present regular wind of $\eta$ Car, and found that they both
exist close to periastron (Kashi \& Soker 2009).
The accretion during the GE was probably an hybrid of Roche lobe overflow and Bondi-Hoyle accretion.
The accreted mass released gravitational energy that might account for the entire
extra energy of the GE, while the jets shaped the bipolar nebula around the star
(the `Homunculus') and supplied part of its kinetic energy.

Damineli (1996) has already tried to connect periastron passages with the light
curve of the GE, and marked the periastron passages on the light curve of the GE,
according to a period of $5.52\yrs$.
He found that the periastron passages fall not far from rises in V-mag.
However, he did not consider the mass loss process, and assumed a constant periodicity,
and for that could not accurately reproduce the times of periastron passages.

In this paper we go back in time to the GE and LE, and examine what were the
binary parameters back then, under our basic assumption that binary companions
trigger major LBV eruptions.
We describe our calculation method and assumptions in section \ref{sec:param}.
In section \ref{sec:results} we present the results of our calculation.
We discuss the implication of our results in section \ref{sec:summary}.

\section{THE HISTORICAL ORBITAL PARAMETERS}
\label{sec:param}

We start with the latest periastron passage in 2009.03
(Corcoran 2009; Fernandez Lajus et al. 2010),
and move back in time to the 19th century, taking a constant orbital period
of $P_0=2023 \days = 5.539 \yrs$ (Damineli et al. 2008a; Fernandez Lajus et al. 2010;
Landes \& Fitzgerald 2009).
The effect of the LE on the binary parameters is small, as only $\lesssim 1 \rm{M_\odot}$
were ejected during this eruption (Smith 2005), and most likely much less then this amount was transferred
from the primary to the secondary.
Nevertheless we take the LE into consideration and check how
it affected the orbital parameters.
Going further back in time we reach the end of the GE, in 1858.
As considerable mass was evidently lost from the system during the GE,
and possibly transferred from the primary to the secondary (Soker 2004),
the binary parameters have changed during the GE.
As mass loss and mass transfer were continuous during the GE,
so are the variations in the binary parameters.

We perform a calculation to determine the change in the orbital parameters,
as described below.
We repeat it twice, one time for the LE and one time for the GE.
The initial conditions for the calculation for the LE are taken from present day
system parameters.
The calculations of the binary evolution during the GE are constrained by the
demand that the binary parameters at the end of the GE are like those at
the beginning of the LE as we derived first.
Practically, we integrate the equations backward in time.

Two major effects are responsible for changing the orbital period and the other
parameters (eccentricity and semimajor axis):
mass loss from the binary system increases the orbital period, while mass transfer
from the primary to the secondary reduces it.
For the binary parameters of $\eta$ Car, per unit mass,
the effect of reducing the orbital period by mass transfer
is much stronger than the effect of increasing it by mass loss.

To calculate the variation in the binary and orbital parameters we
follow the equations from Eggleton (2006).
The rates of change of the stellar masses (going forward in time) are
\begin{equation}
\begin{split}
&\dot M_1=-\dot m_{l1}-\dot m_t ~~;~~
\dot M_2=-\dot m_{l2}+\dot m_t ~~;~~ \\
&\dot M = \dot M_1 + \dot M_2 = -\dot m_{l1} -\dot m_{l2},
\label{eq:Mdot}
\end{split}
\end{equation}
where $\dot m_{l1}$ and $\dot m_{l2}$ are the rates of mass loss to infinity
from the primary and the secondary, respectively,
and $\dot m_t$ is the rate of mass transferred from the primary to the secondary.
The mass lost from the secondary is assumed to come from the transferred mass.
It is ejected by the accretion disk in a bipolar outflow.
For clarity in our following explanations, we shall use the product of
these parameters with the duration of the GE,
$M_{li}=\dot m_{li} t_{\rm{GE}}$ and $M_t=\dot m_t t_{\rm{GE}}$.
To minimize the parameters' freedom, we keep the mass loss rate of each star
and the mass transfer rate constant during each of the 19th century eruptions.

The orbital separation is calculated as a function of time.
The orbital separation $\textbf{\textit{r}}$ varies according to (Eggleton 2006)
\begin{equation}
\ddot{\textbf{\textit{r}}}(t) = -\frac{GM(t)\textbf{\textit{r}}(t)}{r^3(t)} +
\dot m_t \left(\frac{1}{M_1(t)}-\frac{1}{M_2(t)} \right) \dot{\textbf{\textit{r}}}(t).
\label{eq:rt}
\end{equation}
The first term on the right hand side is the unperturbed gravitational acceleration, and
the second term is a perturbing acceleration due to mass transfer and mass loss.
The perturbing acceleration depends linearly on the velocity, and for that
its effect is larger close to periastron passage.
Because we know the orbital period and we have an estimate of the eccentricity at present,
we performed the integration backward in time to just before the GE.
In this procedure the present parameters of $\eta$ Car serve as the initial conditions.
As the changes in the masses are not negligible, the equation cannot be solved
analytically and it is solved numerically, using an automatic step-size
Runge-Kutta-Fehlberg integration method, of order 2-3
(implemented by MATLAB function ODE23, using accuracy factor of $10^{-8}$).

The eccentricity $e(t) \equiv |\textbf{\textit{e}}(t)|$ is calculated according to
\begin{equation}
GM\textbf{\textit{e}} =
\dot{\textbf{\textit{r}}}\times(\textbf{\textit{r}}\times\dot{\textbf{\textit{r}}}) -
\frac{GM\textbf{\textit{r}}}{r}.
\label{eq:e}
\end{equation}
The Keplerian energy per unit reduced mass $\varepsilon(t)$ is calculated according to
\begin{equation}
\varepsilon(t) = \frac{1}{2} \dot{r}^2(t)
- \frac{GM(t)}{r(t)},
\label{eq:energy}
\end{equation}
and then we can calculate the semi-major axis
\begin{equation}
a(t) = - \frac{GM(t)}{2\varepsilon(t)},
\label{eq:a}
\end{equation}
and the orbital period
\begin{equation}
P(t)=2\pi\sqrt{\frac{a^3(t)}{GM(t)}}.
\label{eq:P}
\end{equation}
The specific angular momentum is
\begin{equation}
\textbf{\textit{h}}=\dot{\textbf{\textit{r}}}\times\textbf{\textit{r}}.
\label{eq:h}
\end{equation}

We can summarize our method of calculation and assumptions as follows:
(1) Mass transfer and mass loss rates are constant during the orbital period.
This might not be exactly the case.
Mass loss and mass transfer rates might have been larger
near periastron passage, while the slow relative motion near apastron might have
increased the mass accretion rate near apastron.
(2) Mass transfer and mass loss rates are constant during the entire GE.
This might not be exactly the case as it is likely that these rates decrease toward
the end of the eruption.
(3) The mass is rapidly lost. This is not exactly accurate near periastron passages,
when the secondary orbital motion relative to the primary was larger than the
equatorial wind speed.
(4) The two very rapid rises by $\sim 1.5$~mag during the GE (see below) occurred during
periastron passages.
(5) Present day mass loss values are $ < 10^{-3}~\rm{M_\odot}~\yr^{-1}$
(Hillier et al. 2001; Pittard \& Corcoran 2002),
and therefore have negligible effect on the orbital parameters.
In other words, present day parameters are assumed to be valid at the end of the LE.
(6) The extra energy of the two eruptions comes from mass accretion onto the secondary.
This assumption helps to constrain the binary parameters,
but the main result of this paper, that under the assumption that the peaks of the GE occurred
near periastron passages we can fit the evolution of $\eta$ Car, does not not depend on this assumption.
This will lead us to conclude that very likely all major LBV eruptions are triggered by companions
at periastron passages.
The triggering of a major eruption requires the primary to be in an unstable state according to our model.
That also results in very high mass transfer and mass loss rates.

\section{RESULTS}
\label{sec:results}

\subsection{The Lesser Eruption}
\label{subsec:thelessereruption}

The mass transfer in the LE is $\sim0.04$ times that in the GE,
as the total extra energy was $\sim0.04$ times that in the GE,
if we use our assumption that the extra energy comes from
accretion onto the secondary (see equation \ref{eq:transmass} below).
Under that assumption the accreted mass during the LE is taken to be $0.14 \rm{M_\odot}$.
For the total mass ejected in the LE we take $1\rm{M_\odot}$ (Smith 2005).
The orbital separation as function of time during the LE,
calculated with the prescription described in section \ref{sec:param}, is plotted in Fig. \ref{fig:LE_plot2}.
The parameters used are listed in the figure caption.
The LE did not change the orbital period by much.
For the parameters we used, we find that at the beginning of the LE
(and at the end of the GE) the period was $5.52 \yrs$,
compared with the present period of $5.54 \yrs$.

When going back in time to the LE we find (following Frew 2004) that it began only $\sim 1$~month
after the system was at the 1887.2 periastron.
We assume this is not a coincidence but rather a strong hint that the
periastron passage was the trigger of the LE.
At periastron the secondary is very close to the primary, to the extent
that the tidal force it exerts on the primary is able to reinforce mass transfer and accretion.
We suggest that the periastron passage of 1887.2 occurred when the primary
was in a more unstable phase, e.g. strong magnetic activity (Harpaz \& Soker 2009).
The periastron passage during this more unstable phase triggered the eruption.

One word of caution is in place here.
The light curve of the GE contains strong fluctuations in brightness (Frew 2004).
Namely, some of the peaks might be either observational errors of stochastic variations in $\eta$ Car.
However, the three peaks we based our analysis on (one in the LE and two in the GE)
are much larger than the other peaks, and seem to be real properties of the eruptions.
Though for the two peaks of the GE the error is 0.4-0.5 mag (Frew 2004), it is evident
that the rise was $\ga 1.5~$mag.
The error in the timing in the data by Frew (2004) is small to be of a significance for our results,
which require accuracy of the order of $\sim 0.1 \yr$.

\subsection{Models for the Great Eruption with no mass transfer and no accretion}
\label{subsec:results_nomasstransfer}

As the LE started close to periastron (Frew 2004), we are encouraged to think that a
similar effect has triggered the GE.
The GE has two peaks in V-mag with rise larger than $1.5~$mag, at 1837.9 and 1843.
We use our two basic assumptions to constrain the masses of the two stars.
The two basic assumptions are that (a) major LBV eruptions are triggered by periastron passages,
and (b) that most of the extra energy is released by mass transfer to the non-eruptive secondary.

The GE changed the orbital parameters more dramatically than the LE.
The reasons for that are (a) that more mass was lost to the Homunculus (Smith et al. 2003),
and (b) the much brighter event resulted from higher mass transfer to the secondary according
to one of our assumptions.
We perform the calculation described in section \ref{sec:param} for different sets of parameters.
First, we take the `common model' to be the model with the conventional parameters
$M_1=120 \rm{M_\odot}$ and $M_2=30 \rm{M_\odot}$
(in our previous works we used these stellar masses, adopted from the literature),
and neglect mass transfer.
To be punctilious, the masses we take here are for present day $\eta$ Car,
and are changed by less than $1 \rm{M_\odot}$ during the LE,
so practically they can be referred to as the masses at the end of the GE
as well for this model (which we rule out later).
In the `common model' a mass of $M_{l1} = 12 \rm{M_\odot}$ is ejected by the primary,
and no mass is transferred to the secondary or lost from it: $M_{l2}=M_t=0$.
We find that the orbital period before the GE in this model was $P_{1837.9}=4.76 \yrs$,
and that periastron occurred $\sim 8.5$~months after the event had started.
\begin{figure}[!t]
\resizebox{0.6\textwidth}{!}{\includegraphics{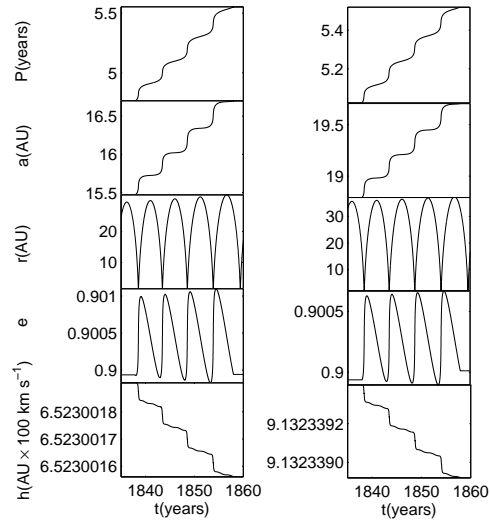}}
\caption{\footnotesize
The variation of the binary parameters
(orbital period $P$, semi-major axis $a$, orbital separation $r$, eccentricity $e$, and specific angular momentum $h$)
during the 20 year long Great Eruption of $\eta$ Car,
for models with no mass transfer from the primary to the secondary.
Left panels: $M=150 \rm{M_\odot}$ (`common model'), right panels: increasing the total mass to $M=250 \rm{M_\odot}$.
Under our assumption that a total mass of $12 \rm{M_\odot}$ lost in a constant rate for the duration of the GE,
the periastron passages do not occur close enough to the two major rises in V-mag,
in 1837.9 and 1843, as has occurred in the LE, for any of the cases with no mass transfer.
Changing the assumption of a constant mass loss rate might improve the fit for
the high mass model, but not for the common model.
}
\label{fig:GE_no_mass_transfer}
\end{figure}

According to one of our assumptions, periastron passages trigger large rapid rises in luminosity.
For that the first orbital period during the GE should be $\sim 5.1 \yrs$
as the time period between the two rapid rises by $\ga 1.5~$mag.
In addition, the peaks should be close to periastron passages, i.e., the orbit should be in phase with luminosity peaks.
If only mass loss from the primary is included, with no mass transfer, then we can fit the period but not the exact times of periastron passage.
In that case the periastron passage from the model occurs after the luminosity rise in 1837.9, which is not the desired outcome.

Under our assumption that the mass loss rate is constant during the GE,
the common model does not match observations.
We could relax the assumption of a constant mass loss rate.
However, more mass loss is expected during periastron passages, a process that enhances
the increase in orbital period due to mass loss.
This will require an even shorter orbital period before the GE,
with a degraded fit to the two luminosity peaks.

Taking a larger mass in the Homunculus, and hence a higher mass loss during the GE,
will also make the discrepancy between the common model and observations larger.
We therefore examine a more massive system for the same mass loss and zero mass transfer.
As evident from equation (\ref{eq:rt}), when $\dot m_t = 0$ the individual masses
of the stars play no role, and the variation in the orbital parameters depends upon the total mass alone.
We find that under the assumption of a constant mass loss rate and a total mass lost during
the GE of $12 \rm{M_\odot}$, a first orbital period of $5.1 \yrs$ years during the GE is achieved
for a pre-GE total mass of $M=(M_1+M_2) \simeq 250 \rm{M_\odot}$ (right panels of Fig. \ref{fig:GE_no_mass_transfer}).
But even for this case, the times of periastron passage do not match the times
of the two rapid rises.
As a criteria for a good fit we take the sum of differences between the times
of periastron passages $P_1$ and $P_2$, and the times of the rises in V-mag, 1837.9 and 1843.0, respectively.
Our criteria for best fit is therefore a minimal value for the quality parameter
\begin{equation}
\Delta P = (P_1-1837.9) + (P_2-1843.0) \yrs.
\label{eq:deltap}
\end{equation}
We find that for the `common model' $\Delta P = 1.12 \yrs$, and for a total system mass of $250 \rm{M_\odot}$
(with no mass transfer) $\Delta P = 0.92 \yrs$.
These values are considerably large comparing to the cases with mass transfer, discussed below.
We therefore conclude that in order to get a good fit for the GE mass transfer must be taken into account,
and that the two stars of $\eta$ Car are much more massive than usually thought.

\subsection{The basic mass accretion model}
\label{subsec:results_basic}

We now turn to use our assumption that the extra energy of the GE, ${E_{\rm GE}}$,
both radiation and kinetic energy, is mainly due to accretion onto the
secondary star (Soker 2001, 2004).
Soker (2007) took a mass of $30 \rm{M_\odot}$ for the secondary star, and deduced that the
accreted mass onto the secondary during the GE was
$M_{\rm{acc}} = M_t-M_{l2} \simeq 8 \rm{M_\odot}$, which is a significant fraction of the secondary mass.
Our finding above that the binary system is much more massive than assumed before,
allows for a much lower total accreted mass onto the secondary, hence accretion caused small
disturbance to the secondary.
We expect, therefore, that the present secondary will have about its main sequence properties.
Taking for the secondary luminosity $L_2 = 9 \times 10^5 \rm{L_\odot}$ (Hillier et al. 2001),
we find its mass to be $M_2\sim 80 \rm{M_\odot}$ (e.g., Meynet \& Maeder 2003).
We can also take for the radius of the secondary, $R_2$, its zero age main sequence (ZAMS) radius, and
constrain the total mass accreted during the GE to
\begin{equation}
\begin{split}
M_{\rm{acc}} &= M_t-M_{l2} \\
&=3.7 \left(\frac{E_{\rm GE}}{8 \times 10^{49} \erg}\right)
\left(\frac{R_2}{14.3 \rm{R_\odot}}\right)
\left(\frac{M_2}{80 \rm{M_\odot}}\right)^{-1}
\rm{M_\odot}.
\label{eq:transmass}
\end{split}
\end{equation}
Similar to Soker (2007), who assumed that $\sim 2/3$ of the material transferred is accreted onto the secondary,
we assume that $M_{\rm{acc}} \simeq 0.65 M_t$, and take
$M_t=5.7 \rm{M_\odot}$ and $M_{l2}=2 \rm{M_\odot}$.
We note that for our basic model `MTz', we use the most basic assumptions,
one of them is that the secondary is a ZAMS star.
This assumption is probably not accurate.
In section \ref{subsec:evolved} below we consider a more realistic model in
which the secondary is an evolved main sequence star, rather than a ZAMS star.

Let us comment on the high mass accretion rate onto the secondary, $\sim 0.2-0.3 \rm{M_\odot} \yr^{-1}$ in our models.
In recent years some other models are based on similar, and even higher, mass accretion rates onto main sequence stars.
In particular we note the merger model for the eruption of V838 Mon.
In the stellar merger model of Tylenda \& Soker (2006) an $8 \rm{M_\odot}$ main sequence star accretes $\sim 0.05 \rm{M_\odot}$ in about 3 months.
This is an accretion rate of $\sim 0.2 \rm{M_\odot} \yr^{-1}$, about equal to the one used here,
but onto a star that is only $\sim 0.1$ times the mass of the secondary here.
Therefore, it seems that the process where main sequence stars accrete at very high rates might occur,
and should be studied in more details.

\begin{figure}[!t]
\resizebox{0.50\textwidth}{!}{\includegraphics{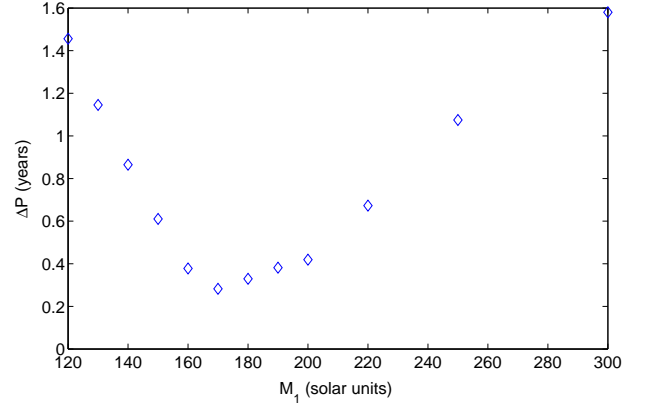}}
\caption{\footnotesize
Fitting the best value for $M_1$ for fixed parameters
$M_2 = 80 \rm{M_\odot}$, $M_{l1}=18 \rm{M_\odot}$, $M_{l2} = 2 \rm{M_\odot}$, and $M_t = 5.7 \rm{M_\odot}$.
$\Delta P$ is our criteria for a good fit, given in equation (\ref{eq:deltap}).
Checking primary masses in the range $M_1 = 120-300 \rm{M_\odot}$ we find best fit for $M_1 \simeq 170 \rm{M_\odot}$,
with $\Delta P = 0.28 \yr$.
By that we determine the value for $M_1$ for our basic model `MTz'.
}
\label{fig:fit_M_1}
\end{figure}

We run a number of models, changing the value of $M_1$,
to find which value best fits the two rises in V-mag, according to our criteria from equation (\ref{eq:deltap}).
The fitting is shown in Fig. \ref{fig:fit_M_1}.
Checking primary masses in the range $M_1 = 120-300 \rm{M_\odot}$ we find best fit for $M_1 \simeq 170 \rm{M_\odot}$,
with $\Delta P = 0.28 \yr$.
By that we determine the value for $M_1$ to complete our basic model `MTz'.
It is interesting to note that the difference between the stellar masses for this case is the same
at $M_1-M_2 = 90 \rm{M_\odot}$ as in the common values used.

With the mass transferred to, and the mass accreted by, the secondary,
with the masses of two stars, and with the orbital period at the end of the
GE fixed, we only need to fix the total mass lost by the primary directly to the
bipolar nebula, i.e., the Homunculus.
We fit $M_{\rm H}=20 \rm{M_\odot}$, well within the estimated range of the Homunculus,
and from that take the mass lost by the primary to the Homunculus to be
$M_{l1}=M_{\rm H}-M_{l2}=18 ~\rm{M_\odot} \yr^{-1}$.
The specific angular momentum is decreasing in a constant rate during the eruptions.
In the common model case where there is no mass loss the
specific angular momentum remains constant (the lower left panel in Fig. \ref{fig:orbital_all});
note that the value of the angular momentum $h$ has a negligible change.
For an interesting discussion of the perturbations to the orbital parameters
in binary systems see Matese \& Whitmire (1984).
For a discussion including the occurrence of non-conservative mass transfer, and its consequences
on parameters of binary systems, see Sepinsky et al. (2009, 2010).

We calculate the orbital parameters of the binary system using these parameters back to the beginning of
the GE using the prescription described in section \ref{sec:param}.
The model with the above parameters is termed `mass transfer ZAMS model' (MTz),
and it is presented in Fig. \ref{fig:orbital_all} and \ref{fig:MTe},
and summarized in Table \ref{Table:comparemodels} together with the other models.

\begin{table*}
\begin{tabular}{||l|l||c|c|c|c|c|c|c||}
\hline \hline
Parameter                       & Symbol                        & Common model & MTz model & MTe model & C1      & C2     & C3     & C4     \\
\hline \hline
Present total mass of the stars & $M (\rm{M_\odot})$            & 150          & 250       &  280      &  200    &  200   &  250   &  250   \\
Present primary mass            & $M_1 (\rm{M_\odot})$          & 120          & 170       &  200      &  120    &  170   &  170   &  170   \\
Present secondary mass          & $M_2 (\rm{M_\odot})$          & 30           & 80        &  80       &  80     &  30    &  80    &  80    \\
Present secondary radius        & $R_2 (\rm{R_\odot})$          & 20           & 14.3      &  23.6     &  14.3   &  14.3  &  14.3  &  14.3  \\
Mass lost by $M_1$ at GE        & $M_{l1} (\rm{M_\odot})$       & 12           & 18        &  38       &  18     &  18    &  12    &  18    \\
Mass lost by $M_2$ at GE        & $M_{l2} (\rm{M_\odot})$       & 0            & 2         &  2        &  2      &  2     &  2     &  2     \\
$M_1$ to $M_2$ Mass transferred & $M_{t} (\rm{M_\odot})$        & 0            & 5.7       &  8.1      &  5.7    &  5.7   &  5.7   &  0     \\
Mass accreted by $M_2$          & $M_{\rm{acc}} (\rm{M_\odot})$ & 0            & 3.7       &  6.1      &  3.7    &  3.7   &  3.7   &  0     \\
Mass in the Homunculus          & $M_{\rm H} (\rm{M_\odot})$    & 12           & 20        &  40       &  20     &  20    &  14    &  20    \\
Orbital period before GE        & $P_{1837.9} (\rm{yrs})$       & 4.765        & 5.370     &  5.239    &  4.990  &  7.702 &  5.574 &  4.734 \\
Present orbital period          & $P_0$ (\rm{yrs})              & 5.539        & 5.539     &  5.539    &  5.539  &  5.539 &  5.539 &  5.539 \\
Present semimajor axis          & $a_0$ (\rm{AU})               & 16.64        & 19.73     &  20.48    &  18.31  &  18.31 &  19.73 &  19.73 \\
Quality of fit                  & $\Delta P$ (\rm{yrs})         & 1.118        & 0.283     &  0.606    &  1.456  &  5.100 &  0.485 &  1.684 \\
\hline \hline
\end{tabular}
\caption{A comparison between the models used in the paper.}
\label{Table:comparemodels}
\end{table*}

We emphasize that `MTz' is not our preferred model.
It simply served as the basic model to which other models will be compared to.
The model `MTz' will serve to check the role of the different
parameters below, and it will serve as the basis to examine more realistic
models that include the expectation that the two stars already evolved from the ZAMS.
In this paper our preferred model is `MTe' (see below),
and we intend to examine more models in future papers.

It is clearly seen from Fig. \ref{fig:MTe} that the two sharp rises by $\ga 1.5~$mag
in luminosity occurred very close to periastron passage.
There is no perfect match, but it should be recalled that we did not perform
too much of a parameter fitting.
For example, we tried to fit masses in increments of only $10 \rm{M_\odot}$,
and only tried two values for the ejected mass.
We only make sure that the scaling of the parameters as described above lead
to an almost perfect fit.
We could make further adjustments to the mass transfer and mass loss by the secondary, as well
as with the stellar masses.
Small variations in the parameters and small changes from a constant
mass loss and mass transfer rates can lead to a perfect match.

The main finding here is that a very massive binary system, where the primary experienced
heavy mass loss and mass transfer, and where the secondary lost part of the transferred mass,
fits the observations very nicely under our assumptions.
The accreted mass onto the secondary can account for the extra energy released during the GE.
Further more, mass transfer and accretion onto the secondary make the fitting
of periastron passages to observed luminosity peaks much easier than do models that
include no mass transfer and accretion.

\subsection{The role of the different parameters}
\label{subsec:results_parameters}

\begin{figure}[!t]
\resizebox{0.6\textwidth}{!}{\includegraphics{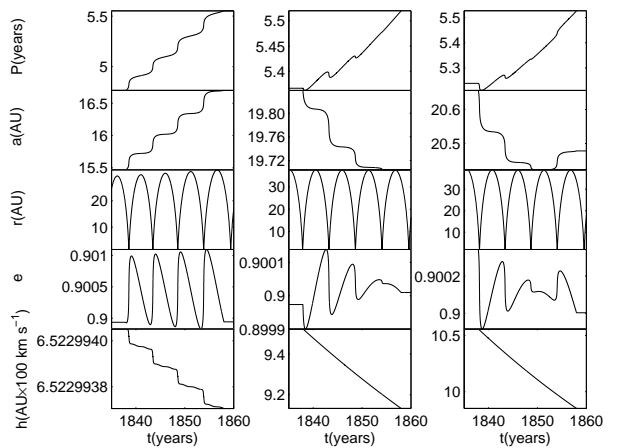}}
\caption{\footnotesize
The variation of the binary parameters
(orbital period $P$, semi-major axis $a$, orbital separation $r$, eccentricity $e$, and specific angular momentum $h$)
during the 20 year long Great Eruption of $\eta$ Car.
The variations are given for different sets of parameters we use in the paper (see Table \ref{Table:comparemodels}).
Left: `Common model', middle: `MTz' model, right: `MTe' model.
}
\label{fig:orbital_all}
\end{figure}

\begin{figure}[!t]
\resizebox{0.50\textwidth}{!}{\includegraphics{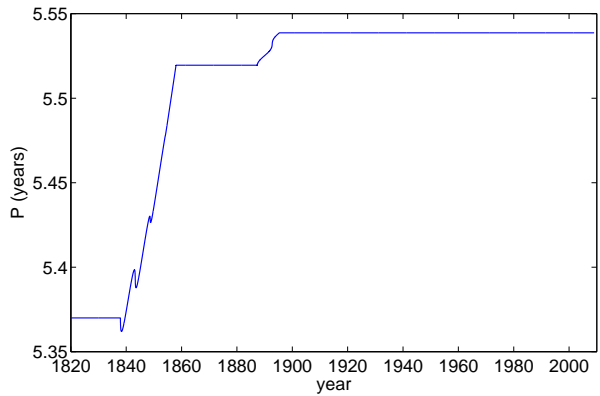}}
\caption{\footnotesize
The historical change of the orbital period for the `mass transfer ZAMS model' (MTz).
The present day orbital period $P_0=2023 \days = 5.539 \yrs$ is valid, when
going back in time up to the end of the LE (1895.3).
During the LE it is slightly reduced, but the more considerable reduction,
still going back in time,occurred during the GE.
The ``$\thicksim$'' like features occur close to periastron passages, and
are more evident during the GE, when the mass transfer rate is higher
than the mass transfer rate during the LE (both assumed to be constants)
(see text for full explanation).
}
\label{fig:P}
\end{figure}

We turn to analyzing the behavior of the orbital parameters,
orbital period $P$, eccentricity $e$, and semi-major axis $a$, during the GE.
Their variation for our basic model `MTz' are presented in the middle column of
Fig. \ref{fig:orbital_all}, together with the orbital separation $r$.
The variation of the orbital period $P$ during both eruptions is presented in Fig. \ref{fig:P}.
We also study the role of the binary masses and amount of transferred and lost masses.
For our basic parameters (`MTz' model), a unit of mass transferred from the primary to
the secondary reduces the orbital period much more than a unit of mass lost from the system
increases it.
Namely, $M_t$ plays a greater role than $M_{l1}$ and $M_{l2}$ in determining the
orbital period.

As the parameters space is large, we restrict ourselves to comparing the quality of the
fit by changing one parameter at a time.
We therefore checked four different sets of parameters, to illustrate their influence on the results.
The four additional models we checked are:
\begin{enumerate}
\item Model `Comparison 1' (`C1'): Decreasing the primary mass to $M_1=120 \rm{M_\odot}$.
This results in a shorter binary period before the GE, as the
mass transfer effect of decreasing the orbital period is smaller
(proportional to $M_1^{-1}-M_2^{-1}$; more negative value implies a larger effect),
and the mass loss effect that increases orbital period is larger.
\item Model `C2': Decreasing the secondary mass to $M_2=30 \rm{M_\odot}$.
This results in a longer binary period before the GE, as the
mass transfer effect of decreasing the orbital period is larger.
\item Model `C3': Decreasing the mass lost by the primary to $M_{l1}=12 \rm{M_\odot}$.
This results in a longer binary period before the GE, as in this model mass transfer that acts
to decrease the orbital period is more important.
\item Model `C4': Decreasing the mass transferred from the primary to the secondary to $M_{\rm{acc}}=0 \rm{M_\odot}$
(namely, no mass transfer, only mass loss).
This results in a shorter binary period before the GE, as no mass transfer implies that the mass
loss effect is more important, and the latter works to increase the orbital period.
\end{enumerate}
Table \ref{Table:comparemodels} gives the parameters for the four cases,
together with the value of the quality of fit parameter, $\Delta P$ (equation \ref{eq:deltap}).
As the parameters space is large, but it is evident that the parameters set for `MTz' gives better
fit then the other sets of parameters, our solution is probably a local minimum for the quality of fit parameter.
Within the possible ranges of parameters, it seems as this is the solution for
the system and mass loss and transfer parameters.

Other parameters, fixed by observations, are the beginning and termination time of the GE.
As the orbit is very eccentric, the number of periastron passages play a major role.
The GE started very close to periastron, and lasted for more than 3 revolutions,
but less than 4.
As according to our model it started shortly after periastron, this means that there were
3 periastron passages during the GE.
The most steep change in the orbital period occurred close to periastron passages.
This is not surprising, since, as we mention above, the perturbing force (per unit mass) in equation \ref{eq:rt},
linearly depends on the orbital velocity, which is considerably larger close to
periastron in such an eccentric orbit.

We note that the value of the $e_0$ itself does not have any effect on the average
change of the orbital period for a whole number of revolutions.
The high eccentricity only causes the period to change at a different rate over the period.
Taking any value of eccentricity does not affect the times of periastron passages.
We only need an eccentric orbit for the secondary to be much closer to the primary
during periastron, so it can trigger the eruption, but any value between
$e_0=0.8$ and $e_0=0.95$ is acceptable for our model.
The upper limit comes from the requirement that the orbital separation at periastron would be
sufficiently smaller than the primary radius.
The primary radius is $\sim 0.83 \AU$ today,
taking primary luminosity of $L_1 \simeq 4.5 \times 10^6 \rm{L_\odot}$ (Davidson \& Humphreys 1997)
and effective temperature of $\sim 20\,000 \K$, and was probably larger during the GE.
The lower limit comes from the requirement that the secondary would be much closer to the primary
during periastron, to result in a significant tidal effect to trigger the eruption.
Note that the same range of eccentricity comes from fitting doppler shifts for lines in present day $\eta$ Car
(Kashi \& Soker 2008b and references therein).
The relative influence of the mass loss and mass transfer varies with orbital phase.
For both processes the change of the orbital parameters per unit time (assuming constant
mass loss and mass transfer rates) near periastron is larger than near apastron.
The difference between periastron and apastron is larger for the mass transfer process.
Therefore, in the first cycles of the GE the mass loss dominates near apastron in increasing
the orbital period, while the mass transfer dominates near periastron and the orbital period
decreases for a short time near periastron (upper middle panel of Fig. \ref{fig:orbital_all}).
This results in the ``$\thicksim$'' like features in the plot of the orbital period vs. time (Fig. \ref{fig:P}).
As evident from the middle column of Fig. \ref{fig:orbital_all}, the magnitudes of the orbital period and semimajor axis
decrease near periastron diminish over the cycles.
This is explained as follows.
The effect of the mass loss process on increasing the orbital period is proportional to $(M_1+M_2)^{-1}$,
while that of the mass transfer in decreasing it is proportional to $M_1^{-1}-M_2^{-1}$.
As the total mass decreases during the GE the effect of the mass loss increases,
while the effect of mass transfer decreases because $M_1$ decreases and $M_2$ increases.
This causes the ``$\thicksim$'' like features which occur close to periastron passages
to weaken from periastron to periastron, up to disappearing in the last one.

The different rates in the changes (sometimes of opposite sign) of the orbital parameters between
periastron and apastron make an exact fitting difficult.
The reason is that fitting is sensitive to whether the GE ended near periastron or near apastron.
Namely, the fit is sensitive to the way the mass loss rate and mass transfer rate
declined at the end of the GE.
For example, a whole number of periods (for constant mass transfer and loss rates)
would lead to no change in the eccentricity.
However, the number of periods during the GE was $\sim 3.5$,
and the eccentricity changed, but by a tiny amount for the `MTz' model.
It varies by the maximum value of $\delta e = 5 \times 10^{-4}$ relative
to its present value $e_0=0.9$.
However, having a large eccentricity the semi-major axis acquired
most of its change in jumps close to periastron passages.

If the GE had terminated earlier (later), the final orbital period would
have been shorter (longer).
The eccentricity, however is less sensitive to the exact termination time of
the GE for the parameters we used, as its amplitude of change is small compared
to its value.
As we divide the mass loss and transfer in the duration of
the GE, the duration only determine their rate, but not how much
in total mass was lost or transferred.
According to our model `MTz' the secondary was in apastron at 1856.88,
and the following periastron was in 1859.63.
The exact end date of the GE is not very important as long as it does
not include another periastron passage, namely, as long it is before $\sim 1859.5$.
As the end date of the GE is commonly considered to be as late as 1856-8, we conclude
that the exact end date has a minor effect on our results.

We conclude that it is not possible, under our assumptions,
to use a low mass system of $\sim 150 \rm{M_\odot}$ to fit the rises in the light curve of the GE.
Also, for our basic model `MTz' with a total binary mass of $250 \rm{M_\odot}$,
we find that mass transfer from the primary to the secondary is a necessary process to fit these rises.

\subsection{Our preferred model: an evolved secondary star}
\label{subsec:evolved}

The model `MTz' serves as the basis model,
for comparison with more realistic models.
In model `MTz' (section \ref{subsec:results_basic}) the secondary star is
taken to be a ZAMS star with a radius of $14.3 \rm{R_\odot}$.
We now consider the evolved nature of $\eta$ Car, and consider an evolved secondary star,
as we know the primary is already in its LBV stage of evolution and must be evolved.
Based on the results of Verner et al. (2005) we take the secondary radius to be
$23.6 \rm{R_\odot}$, but keep its post-GE mass at $M_2=80 \rm{M_\odot}$.
To account for the energy released in the accretion onto the secondary, the accreted mass should be
$M_{\rm{acc}} = 6.1 \rm{M_\odot}$ (\ref{eq:transmass}).
This value is more in line with the values suggested by Soker (2007).

As mass transfer acts to increase the orbital period, a higher accreted and
transferred masses should be compensated for by a change in other parameters, such that an
increase in the primary mass and in the total mass lost to the Homunculus.
We make two additional changes in the evolved `MTe' model, from the basic `MTz' model.
(1) We take more mass to be lost to the Homunculus, as implied by the
results of Smith \& Ferland (2007) and Gomez et al. (2009).
We emphasize that we do not know the Homunculus mass, but rather fit it (but our calculations favor a
large Homunculus mass).
(2) We increase the primary mass.
The increase in the primary mass is required in light of the very massive Homunculus in this model.
As we see, the pre-GE mass of the primary is $\sim 250 \rm{M_\odot}$.
Therefore, in our preferred model, $\eta$ Car turns out to be a very massive binary system.

\begin{figure}[!t]
\resizebox{0.50\textwidth}{!}{\includegraphics{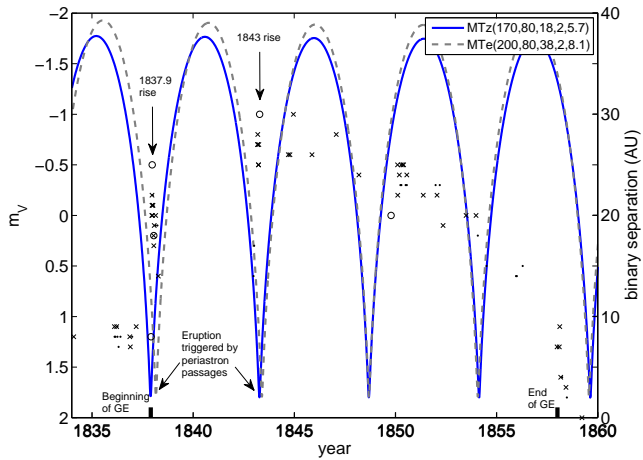}}
\caption{\footnotesize
The binary separation (right axis; for two models) and the V-mag light curve
(left axis; data from Frew 2004, see legend in Fig. \ref{fig:historic_eta_car})
during the Great Eruption of $\eta$ Car for the `mass transfer ZAMS model' (MTz)
(solid blue line), and the `MTe' model (dashed gray line).
The parameters are given in the legend in the order
$(M_1,M_2,M_{l1},M_{l2},M_t)$;
see Table \ref{Table:comparemodels} for definitions of symbols.
`MTe' is our preferred model, taking into account that the two stars already evolved from the ZAMS.
For our preferred model, as well as the `MTz model', two periastron passages nicely fit the
rapid rises in V-mag in 1837.9 and 1843.
}
\label{fig:MTe}
\end{figure}

As before, we do not look for a perfect match, and change $M_1$ by increments of $10 \rm{M_\odot}$.
We find a good fit of the orbital passages to the GE peaks, as shown in Fig. \ref{fig:MTe},
for the following parameters:
A primary and a secondary post-GE masses of $M_1=200\rm{M_\odot}$ and $M_2=80\rm{M_\odot}$, respectively,
where the primary has lost $M_{l1}=38 \rm{M_\odot}$ directly to the
Homunculus and transferred
$M_t=8.1 \rm{M_\odot}$ to the secondary, out of which the secondary accreted
$M_{\rm{acc}}=6.1 \rm{M_\odot}$
(equation \ref{eq:transmass}), and lost the rest $M_{l2}=2 \rm{M_\odot}$.

The results of this section strengthen those of the previous section.
Models with an evolved massive binary system, a massive Homunculus, and
where the extra energy of the GE is supplied in a large part by accretion onto the
secondary star, can account for the two peaks of the GE.

\section{SUMMARY AND DISCUSSION}
\label{sec:summary}

The Lesser Eruption (LE) of $\eta$ Car started close to
the periastron passage of 1887.3 (Frew 2004), as evident from Fig.
\ref{fig:LE_plot2}.
This finding suggests that the interaction with the secondary,
the less massive star, near periastron passage triggered the major eruption.
We generalize this conclusion, and assume that the two rises in luminosity
of the Great Eruption (GE) in 1837.9 and 1843 were also triggered by periastron passages.
In addition to the strong tidal interaction exerted on the primary near periastron
passage, major eruptions such as the LE and GE require the primary to be in
a very unstable phase.

In addition to the assumption that the GE peaks in the light curve occurred
near periastron passages, we assume for simplicity that mass was lost from the
system and that mass was transferred to the secondary at constant rates
during the GE (to make parameters fitting minimal),
that the total mass in the bipolar nebula (the Homunculus) is $12-40 \rm{M_\odot}$
(Smith 2005; Smith \& Ferland 2007; Gomez et al. 2009), and that the mass
transferred onto the secondary released gravitational energy that can account for
most of the extra energy of the GE (Soker 2007).
Not only the energy budget, but also the shape of the light curve, supports
a mass transfer source for the extra energy.
We note, though, that the assumption of accretion is not a necessary one for the
results of the present paper, and it mainly used to constrain the amount of
mass transferred from the primary to the secondary during the GE.

The 1837.9 peak increased sharply and $\sim 4$~months later declined.
A single star model is unlikely to produce this very narrow first peak, as
typical pulsation of unstable star do not have this type of behavior.
Accretion onto the secondary, on the other hand, can easily account for that
because the duration and shape can be accounted for by an accretion event (Soker 2007).
After the first peak, the high and continuous mass loss rate from the LBV
initiated close to the next periastron passage of 1843.
The following two periastron passages may have prolonged the eruption,
possibly by extending the period of high mass loss rate.

Solving for the orbital period evolution as a result of mass loss by
the two stars and mass transfer from the primary to the secondary,
which effectively gives a shorter orbital period before the GE (section \ref{sec:results}),
we could fit the 1837.9 and 1843 sharp rises in the V-mag light curve with
the occurrence of periastron passages (Fig. \ref{fig:MTe}).
However, in order to achieve that fit we had to use stellar masses much larger
than the previously used values.

We used two models.
Model `MTz' assumes (wrongly) that the secondary is not
evolved, and we took it to be a ZAMS star.
This is our basic model; other models here are compared to
it, as well as models we intend to run in future papers.
In our preferred and more realistic model `MTe', the secondary is evolved,
with a radius and temperature as in Verner et al. (2005).
We find that a very good fit for the motel `MTz' is obtained for post-GE
stellar masses of $M_1 \simeq 170\rm{M_\odot}$, $M_2\simeq80\rm{M_\odot}$,
mass of $20\rm{M_\odot}$ lost in the GE,
and $5.7\rm{M_\odot}$ transferred from the primary to the secondary.
For model `MTe' a good fit is obtained for
$M_1 \simeq 200\rm{M_\odot}$, $M_1\simeq80\rm{M_\odot}$, with $40\rm{M_\odot}$ lost in the GE,
and $8.1\rm{M_\odot}$ transferred from the primary to the secondary.
A detailed comparison between the parameters of the models is given in Table \ref{Table:comparemodels}.

In the calculations we neglected tidal friction as a result of tides the two stars
rise on each other close to periastron,
as well as the drag force from the surrounding ejecta that is not accreted.
Both effects act to reduce the orbital period.
However, as we estimate below, their effect is not expected to be large,
although it can be non-negligible in the sense that the required transferred mass
might be reduced by $\sim 5-30 \%$ if these effects are included.
The drag force due to particles that are not accreted is given by Alexander et al. (1976),
and we used it as it was used by Livio \& Soker (1984)
\begin{equation}
F_d = \pi \ln \left(\frac{R_{\rm{max}}}{R_{\rm{BH}}}\right) R_{\rm{BH}}^2 v_{\rm{wind1}}^2
\left(1-\frac{L_{\rm{2,e}}}{L_{\rm{2,Edd}}}\right)^2,
\label{eq:dragforce}
\end{equation}
where $R_{\rm{BH}}$ is the Bondi-Hoyle accretion radius, $R_{\rm{max}}$ is the cut-off distance,
$v_{\rm{wind1}}$ is the relative velocity between the secondary and the primary wind,
$L_{\rm{2,e}}$ is the secondary luminosity, and $L_{\rm{2,Edd}}$ is the secondary Eddington luminosity.
It is assumed that gas particles within a radius $R_{\rm{BH}}$ are accreted by the secondary.
The drag force has the role of $\dot M_{\rm{\rm acc}} v_{\rm{wind1}}$ that is played by the accreted mass
in reducing the orbital separation.

The large uncertainty is in the cut-off distance $R_{\rm{max}}$.
This is the distance beyond which the gravity of the secondary is not the main force
that determines the trajectory of the gas.
Traditionally, $R_{\rm{max}}$ is taken to be the mean radius of the secondary's potential lobe.
However, it seems that during the GE $R_{\rm{max}}$ was small, and the drag force weak.
The cause for that was the huge luminosity of the system, whether it comes from the primary or the secondary.

More specifically, in our model the secondary has a super-Eddington luminosity,
explained by accretion through the equatorial plane and an accretion disk.
Most of the radiation escapes in a wide angle over the secondary's polar directions.
Indeed, most of the mass of the Homunculus is in two polar caps (Smith 2006).
Therefore, the mass that was lost by the primary and passed near the secondary,
was mainly flowing in the equatorial plane, and was accreted.
Most of the influence on the orbit occurs near periastron, namely, at small radii.
The high accretion rate implies that the effective accretion radius ($R_{\rm{BH}}$) was very large.
Had the material been left around the secondary, it would have been able to influence its orbit by
exerting tidal forces.
However, as the accretion radius is large the material was not left around the secondary but rather accreted onto it,
and therefore it was not left there to exert the tidal forces.

During the GE the binary system was within an optically thick region,
and the luminosity was super-Eddington.
This implies that radiation pressure on the gas in the polar directions was significant.
The radiation and jets that were launched by the accreting secondary were therefore
effective in pushing the gas in the polar directions,
further reducing the amount of gas that is influenced by the gravity of the secondary.
Over all, we estimate that $R_{\rm{max}}$ was not much larger than $R_{\rm{BH}}$, and that
$\ln \left(\frac{R_{\rm{max}}}{R_{\rm{BH}}}\right)<1$.

We also note that it is impossible to replace accretion by drag alone.
The gravity of the secondary must influence the flow up to a large distance, in order to make
the drag force significant without accretion taking place.
In that case we would expect accretion to occur.
It seems one cannot escape the conclusion that significant accretion must have occurred during the GE.

For reasonable parameters under our assumptions, we constrained the stellar masses
to post-GE values of $M_1 \simeq 150-200 \rm{M_\odot}$, and $M_2 \simeq 70-90 \rm{M_\odot}$.
We note that a secondary with a ZAMS mass of $M_2 \simeq 70-85 \rm{M_\odot}$ will have the
required luminosity we use here of $L_2 \simeq 9 \times 10^5 \rm{L_\odot}$ (Verner et al. 2005).
This is also the upper luminosity limit in the regime plotted for the secondary in the H-R diagram
(Mehner et al. 2010), though this authors found a lower favorable value of $L_2 \simeq 4 \times 10^5 \rm{L_\odot}$.
For a primary luminosity of $L_1 \simeq 4.5 \times 10^6 \rm{L_\odot}$ (Davidson \& Humphreys 1997),
a ZAMS mass $M_1 \simeq 230 \rm{M_\odot}$ is required (Figer et al. 1998).
We note that Figer et al. (1998) considered for the Pistol star a mass of
$\sim 200-250 \rm{M_\odot}$ and a metallicity slightly above solar metallicity.
It seems as if very massive stars with solar metallicity can be formed in our Galaxy.
Figer (2005) posed an upper limit for the most massive single star at $150 \rm{M_\odot}$.
One possibility for forming such massive primary is a coalescence of two very massive stars,
in a collision or a merger process.
A similar process was discussed by Tutukov \& Fedorova (2008) as a possible mechanism for forming
massive blue stars.
The high rotational velocity of the primary, $\sim0.8-0.9$ of critical velocity
(van Boekel, et al. 2003; Aerts et al. 2004; Smith et al. 2004) is also supported by this scenario.
It is interesting to note that recently Crowther et al. (2010)
observed a few very massive stars in the R136 star cluster,
with the most massive member having presently a mass of $\sim 265 \rm{M_\odot}$,
but possibly started as a $320\rm{M_\odot}$ ZMAS star.

The largest uncertainty is the variation of mass loss and mass transfer during the orbit.
The effects of both mass loss and mass transfer are much larger if occur near periastron passages.
As we have not much information, we simply used a constant values during the GE.
If, for example, tidal interaction near periastron passages caused burst of mass ejection by
the primary, but the accretion onto the secondary continues for the entire orbit, then
we can do with less total mass loss and less massive primary.
A resembling process occurs in cataclysmic variables, where mass transfer rate depends
on the shape of the Roche lobe of the donor, and sensitive to many other parameter,
resulting in a varying mass loss rate (Ritter 1988).
Because of the large uncertainties in the variations along the orbit of the
mass loss and mass the transfer rates, we presently cannot strongly constrain the stellar masses.
We can only say with a high confidence that before the GE the total
mass was $M_1+ M_2 \ga 250 \rm{M_\odot}$.

With the parameters from the GE for model `MTz' at hand, we reexamined the fitting of the LE.
We could match the beginning of the LE with a periastron passage
(solid blue line in Fig. \ref{fig:LE_plot2}).
The same fitting was performed for model `MTe'.
This shows that the LE is not sensitive to the uncertainties of the mass
transfer and mass loss during the GE, and that our assumption that periastron passages trigger
major eruptions is a reasonable assumption.

The reason we find a high secondary mass comes directly from our assumption that the luminosity
of the GE was a result of mass accretion.
For that, the larger the mass of the secondary, the lower the required accreted mass and vice versa (equation \ref{eq:transmass}).
Mass transfer works to decrease the orbital period, while mass loss works to increase it.
If we take the mass transfer to be too large it will not allow a good fit to the binary period
we assume existed at the beginning of the GE (with periastron passages close to 1837.9 and 1843.0).
For that our model requires that the mass of the secondary be larger than the value usually assumed,
in order to account for the GE luminosity.

A very massive binary systems is supported by, or compatible with, other
considerations:
\begin{enumerate}
\item Traditionally the masses of the two stars are determined by the assumption
that each star shines at its Eddington luminosity limit (Davidson \& Humphreys 1997).
This assumption, however, is not in agreement when considering the
evolutionary paths of main sequence stars.
Instead, we took the approach that the stars possess their luminosity as given
by stellar evolution schemes.
For a secondary stellar luminosity of $L_2 \simeq 9 \times 10^5
\rm{\rm{L_\odot}}$ (Verner et al. 2005), the MS secondary mass should be
$M_2 \simeq 85 \rm{M_\odot}$ (Meynet \& Maeder 2003).
The stellar evolution calculations for very massive stars conducted by
Yungelson et al. (2008; see also Figer et al. 1998) show that a star with a
main sequence mass of somewhat more than $200 \rm{M_\odot}$ would become an LBV
with the same luminosity of $L_1 \simeq 4.5 \times 10^6 ~\rm{L_\odot}$ and effective temperature
$T_1 \simeq 20,000 \K$ as that thought for the primary of $\eta$ Car (Verner et al. 2005).
\item A more massive primary would make it easier to account
for the super-Eddington luminosity during the GE, $L_{\rm GE} \simeq 2\times 10^7 ~\rm{L_\odot}$,
because the ratio $L_{\rm GE}/L_{\rm{Edd}}$ is smaller.
\item In Kashi \& Soker (2009) we showed that the massive binary system model
(termed there `high-masses model'), favors accretion onto the secondary close
to periastron passage in present time.
Such an accretion can explain the termination of the secondary wind that is
required (Soker 2005a; Kashi \& Soker 2009) to account for the deep decline in
the X-ray emission (Corcoran et al. 2001; Corcoran 2005).
\end{enumerate}

We speculate that major LBV eruptions (also termed `giant eruptions'; Smith \& Owocki 2006),
are the result of unstable LBVs that are perturbed by the interaction
with a companion during a periastron passage.
This does not happen every periastron passage because the primary LBV must enter
an unstable phase for that to occur.
Our suggestion requires further study before it can be confirmed.
Very few detailed major LBV eruptions have been observed.
More observations of LBV binaries are a necessity to study this type of eruptions.
For example, the series of major eruptions of the LBV in NGC 3432 (Pastorello et al. 2010), which is
possibly also a binary system, may be used to learn more about these eruptions.

Because of the relatively detailed documentation of the GE,
it serves as a prototype for major LBV eruption.
We suggest that the LE of $\eta$ Car and the 17th century eruption of P~Cygni
(Kashi 2010) were also triggered by a periastron passage of the companion during
a very unstable phase of the primary star.
In extreme cases the companion can accrete a considerable amount of mass
and liberate a huge amount of gravitational energy during a short time,
forming an optical transient event.
In Kashi et al. (2010) we suggest that a somewhat similar process induced
the optical transient NGC~300~2008OT-1 (observed by Monard 2008; Bond et al. 2009),
and can account for other optical transients as well.

According to our calculation, and under our assumptions $\eta$ Car is the most massive binary system
in the galaxy, leaving behind other massive stars, such as the Pistol star
(Figer et al. 2004; Najarro 2005), WR 102ka (Barniske et al. 2008)
and LBV 1806-20 (Figer et al. 1998).
However, we also predict that all other massive stars that went through a
non-spherical high mass loss rate phase have massive companions, e.g., P~Cygni (Kashi 2010).
Nota et al. (1995) already suggested that the non-spherical nebulae
around LBV stars are shaped by binary interaction.
So, it might still be possible for the Pistol star, for example, to take the lead.

\acknowledgements
We thank Peter Eggleton, Alceste Bonanos, Orsola de Marco, Mario Livio, Nathan Smith ,Otmar Stahl
and an anonymous referee for very valuable comments that improved the paper.
We especially thank Peter Eggleton for pointing out the necessity to include a model
of an evolved secondary star, and for many suggestions that improved the paper.
We acknowledge Eduardo Fernandez Lajus for using his data file of the historical
observations of Eta Carinae.
This research was supported by the Asher Fund for Space Research at the
Technion, and the Israel Science Foundation.

\end{document}